\documentclass{PoS}

\title{Monopole Dominance of Confinement in SU(3) Lattice QCD}

\ShortTitle{Monopole Dominance of Confinement in SU(3) Lattice QCD}

\author{\speaker{Hideo Suganuma}, \\
        Department of Physics \& Division of Physics and Astronomy, 
Graduate School of Science, \\
Kyoto University, 
Kitashirakawaoiwake, Sakyo, Kyoto 606-8502, Japan\\
        E-mail: \email{suganuma@scphys.kyoto-u.ac.jp}}

\author{Naoyuki Sakumichi \\
Ochanomizu University, 
2-1-1 Otsuka, Bunkyo, Tokyo 112-8610, Japan}

\abstract{
To check the dual superconductor picture for the quark-confinement mechanism, we evaluate monopole dominance as well as Abelian dominance of quark confinement 
for both quark-antiquark (Q$\bar{\rm Q}$) and three-quark (3Q) systems 
in SU(3) quenched lattice QCD in the maximally Abelian (MA) gauge. 
First, we examine Abelian dominance for the static Q$\bar{\rm Q}$ system 
in lattice QCD with various spacing $a$ at $\beta=5.8-6.4$ and various size $L^3 \times L_t$. 
For large physical-volume lattices with $La \ge 2{\rm fm}$, 
we find {\it perfect Abelian dominance} of the string tension for the Q$\bar{\rm Q}$ systems: $\sigma_{\rm Abel} \simeq \sigma$.
Second, we accurately measure the static 3Q potential for more than 300 different patterns of 3Q systems with 1000-2000 gauge configurations 
using two large physical-volume lattices: $(\beta, L^3 \times L_t)$=(5.8, $16^3 \times 32$) and (6.0, $20^3\times 32$).
For all the distances, the static 3Q potential is found to be well described by the Y-Ansatz, i.e., two-body Coulomb term plus three-body
Y-type linear term $\sigma L_{\rm min}$, where $L_{\rm min}$ is the minimum flux-tube length connecting the three quarks.
We find {\it perfect Abelian dominance} of the string tension also for the 3Q systems: $\sigma^{\rm Abel}_{\rm 3Q} \simeq \sigma_{\rm 3Q} \simeq \sigma$.
Finally, we accurately investigate monopole dominance in SU(3) lattice QCD  
at $\beta$=5.8 on $16^3 \times 32$ with 2,000 gauge configurations. 
Abelian-projected QCD in the MA gauge has not only the color-electric current $j^\mu$
but also the color-magnetic monopole current $k^\mu$, which topologically appears.
By the Hodge decomposition, the Abelian-projected QCD system  
can be divided into the monopole part ($k_\mu \ne 0$, $j_\mu=0$) 
and the photon part ($j_\mu \ne 0$, $k_\mu=0$). 
We find monopole dominance of the string tension for Q$\bar{\rm Q}$ and 3Q
systems: $\sigma_{\rm Mo} \simeq 0.92 \sigma$.
While the photon part has almost no confining force, 
the monopole part almost keeps the confining force.  
}

\FullConference{The XIIIth Quark Confinement and the Hadron Spectrum, \\
                  31 July - 6 August 2018\\
                 Maynooth University, Maynooth, Ireland}

\begin{document}

\section{Introduction: Dual Superconductor Picture and Maximally Abelian Gauge}

Quantum chromodynamics (QCD) is the fundamental theory of the strong interaction, 
but the QCD system is highly complicated and is still unsolved analytically 
because of its strong coupling in the low-energy region. 
In particular, quark confinement is an outstanding strange phenomenon 
exhibited in nonperturbative QCD, 
because the fundamental degrees of freedom, quarks and gluons, cannot be observed,
and there is almost no similar phenomenon in other region of physics.
In fact, to clarify the confinement mechanism is one of the most difficult important 
unsolved problems remaining in modern physics.

For the quark-confinement mechanism, 
Nambu, 't~Hooft and Mandelstam proposed 
a dual-superconductor picture in 1970's \cite{N74}. 
In this picture, the QCD vacuum is regarded as 
a color-magnetic monopole condensed system, 
and the dual Meissner effect forces 
the color-electric flux between (anti)quarks 
to be squeezed into one dimension,  
which leads to the flux-tube picture of hadrons \cite{tH81,EI82}.
However, there are two large gaps between QCD and the 
dual-superconductor picture \cite{IS99}.
\begin{enumerate}
\item
The dual-superconductor picture is based on the Abelian gauge theory 
subject to the Maxwell-type equations, 
but QCD is a non-Abelian gauge theory.
\item
The dual-superconductor picture needs 
color-magnetic monopole condensation as the key concept, 
but QCD does not have such a monopole as the elementary degrees of freedom.
\end{enumerate}
As a possible connection from QCD to the dual superconductor picture, 
't~Hooft proposed ``Abelian projection'' \cite{tH81,EI82} 
as an infrared Abelianization scheme of QCD. 
In the Abelian projection, magnetic monopoles topologically appear, and 
't~Hooft conjectured that long-distance physics like confinement 
is realized only by Abelian degrees of freedom in QCD \cite{tH81}, 
which is called ``(infrared) Abelian dominance''.

Actually, in the maximally Abelian (MA) gauge \cite{KSW87}, 
off-diagonal gluons acquire a large effective mass of about 1GeV \cite{AS99}, 
which makes infrared QCD Abelian-like \cite{SY90}, 
and lattice QCD shows appearance of a large clustering of 
the monopole current covering the four-dimensional space-time \cite{KSW87,SNW94}.
In fact, infrared QCD in the MA gauge seems to behave as an Abelian dual-superconductor.

In SU(3) lattice QCD, MA gauge fixing \cite{SS14,SS15} is performed by maximizing 
\begin{eqnarray}
R_{\rm MA}[U_\mu(s)]
\equiv \sum_{s} \sum_{\mu=1}^4  
{\rm tr}\left( U_\mu^\dagger(s)\vec H U_\mu(s)\vec H\right)  
= \frac{1}{2} \sum_{s} \sum_{\mu=1}^4  
\left( \sum_{i=1}^3 |U_\mu(s)_{ii} |^2 -1 \right), 
\end{eqnarray}
under SU(3) gauge transformation.
Here, $U_\mu (s)$ is the link-variable 
$U_\mu (s)=e^{iagA_\mu(s)} \in {\rm SU(3)}_c$ 
with lattice spacing $a$, gauge coupling $g$ and gluon fields $A_\mu$. 

The Abelian link-variable 
$u_\mu(s) = e^{i\theta_\mu(s) }=e^{i\theta_\mu^3(s) T_3 + i\theta_\mu^8(s) T_8}
\in {\rm U(1)}^2$
is extracted from the link-variable
$U_\mu^{\rm MA} (s) \in {\rm SU(3)}_c$ in the MA gauge \cite{SS15}, 
by maximizing the overlap of $R_{\rm Abel} \equiv \frac{1}{3}
{\rm Re} \, {\rm tr}\left( U_\mu^{\rm MA}(s) u_\mu^\dagger(s) \right) 
\in [-\frac{1}{2},1]$. 
Maximally Abelian projection is defined by 
the replacement of $\{U_\mu^{\rm MA}(s)\} \rightarrow \{u_\mu(s)\}$, 
which corresponds to the elimination of off-diagonal gluon components.

In this paper, to check the dual superconductor picture, 
we accurately investigate Abelian dominance  \cite{SS14,SS15} and monopole dominance 
of the quark confining force for both quark-antiquark 
(Q$\bar{\rm Q}$) and three-quark (3Q) systems in SU(3) quenched lattice QCD 
in the MA gauge. 
For the error estimate, we use the jackknife method.

\section{Perfect Abelian dominance of quark confinement 
in quark-antiquark systems}

First, we study the static Q$\bar{\rm Q}$ potential $V(r)$ 
in lattice QCD with 
$(\beta, L^3 \times L_t)=(6.4, 32^4), (6.0, 32^4)$ and $(5.8, 16^3 \times 32)$ 
\cite{SS14,SS15}.
The Q$\bar{\rm Q}$ potential $V(r)$ 
is obtained with the Wilson loop $W_{r \times t}[U_\mu]$, 
and its MA projection (Abelian part) $V_{\rm Abel}(r)$ 
is similarly defined by the Abelian Wilson loop $W_{r \times t}[u_\mu]$,
\begin{eqnarray}
V(r)=-\lim_{t \rightarrow \infty} 
\frac{1}{t}{\rm ln} \langle W_{r \times t}[U_\mu] \rangle, \qquad
V_{\rm Abel}(r)=-\lim_{t \rightarrow \infty} \frac{1}{t}{\rm ln} 
\langle W_{r \times t}[u_\mu] \rangle.
\end{eqnarray}

We show in Fig.1(a) the Q$\bar{\rm Q}$ potential $V(r)$ 
and its Abelian part $V_{\rm Abel}(r)$. 
They are found to be well reproduced by the Coulomb-plus-linear Ansatz, respectively:
\begin{eqnarray} 
V(r) = - \frac{A}{r} + \sigma r + C,  
\qquad
V_{\rm Abel}(r) 
= - \frac{A_{\rm Abel}}{r} + \sigma_{\rm Abel}~r + C_{\rm Abel}. 
\end{eqnarray}
Figure~1(b) shows the difference $V(r)-V_{\rm Abel}(r)$ plotted with $r$ at each lattice. 
As a remarkable fact from Fig.1, we find {\it perfect Abelian dominance} 
of the string tension, $\sigma_{\rm Abel} \simeq \sigma$, for the Q$\bar{\rm Q}$ system.

\begin{figure}[h]
\centering
\includegraphics[height=5cm]{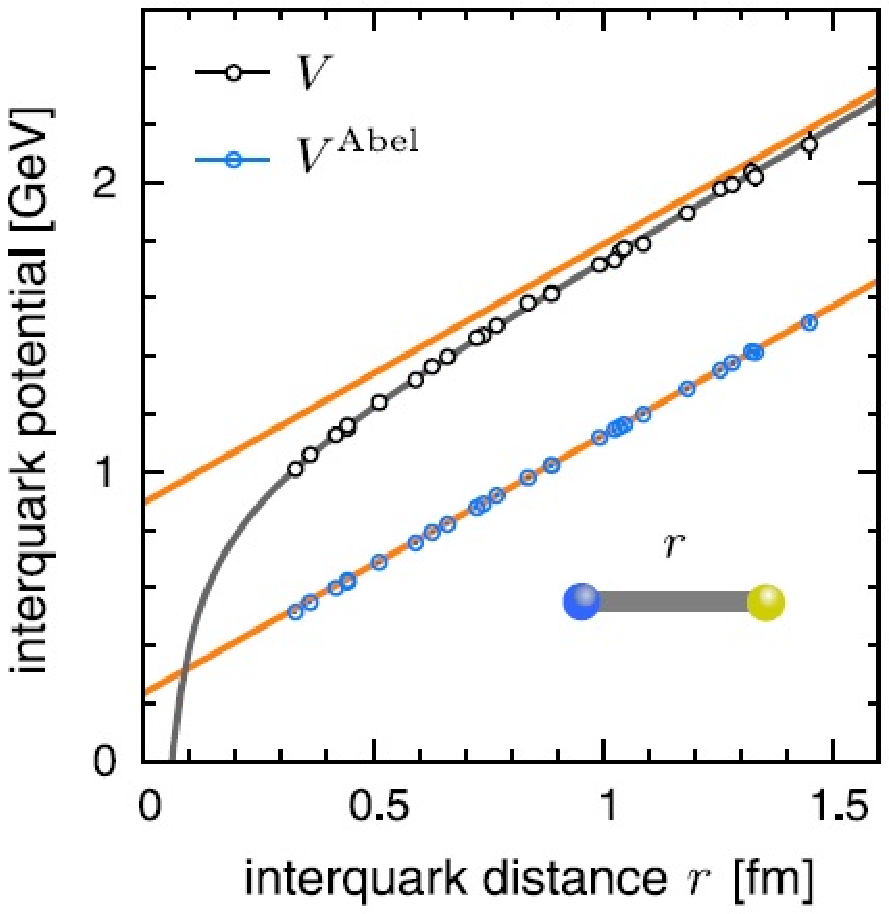}
\hspace{0.5cm}
\includegraphics[height=5cm]{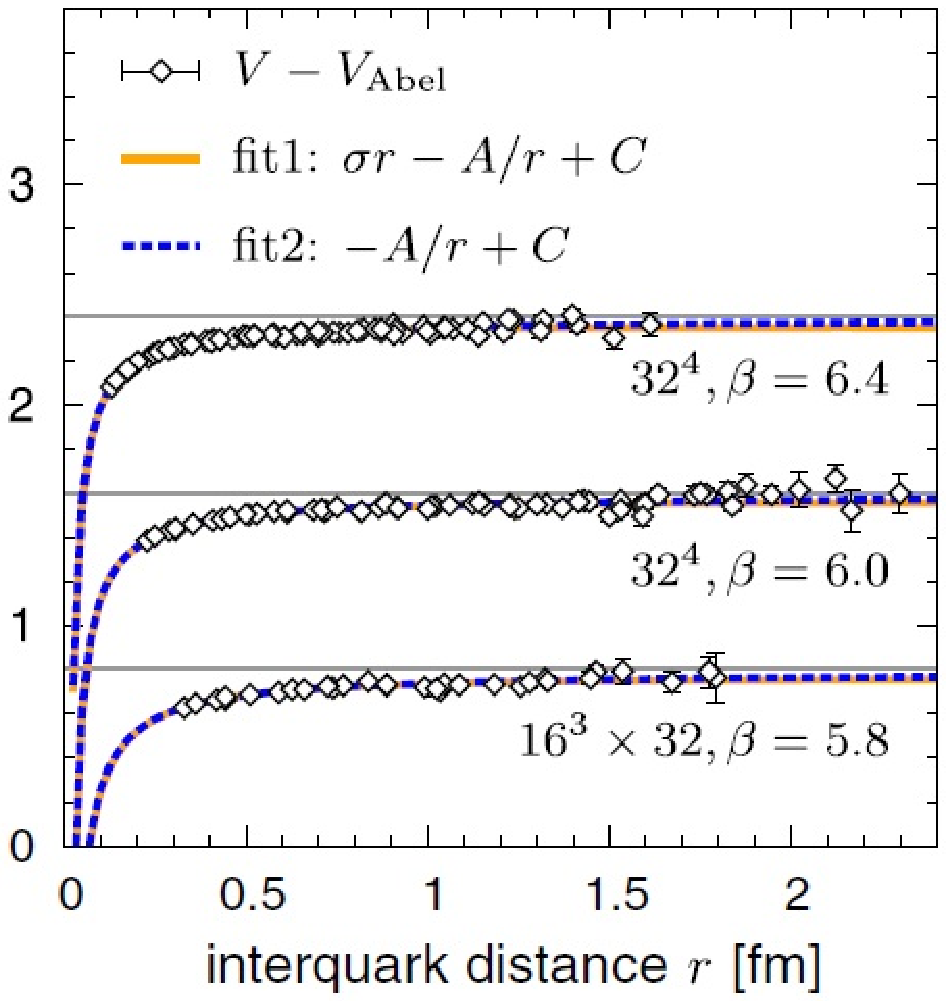}
\caption{
(a) The lattice QCD result of the Q$\bar{\rm Q}$ potential $V(r)$ (black) and 
its Abelian part $V_{\rm Abel}(r)$ (blue) for $(\beta, L^3 L_t)=(5.8, 16^3 32)$ 
\cite{SS15}.
(b) $V(r) - V_{\rm Abel}(r)$ for 
$(\beta, L^3 L_t)=(6.4, 32^4), (6.0, 32^4)$ and $(5.8, 16^3 32)$ \cite{SS14}. 
At each lattice, all the data can be well fit with the pure Coulomb form 
with $\sigma=0$, which means $\sigma_{\rm Abel} \simeq \sigma$.
}
\end{figure}

We also examine the physical lattice-volume dependence 
of $\sigma_{\rm Abel}/\sigma$ in Fig.2, and find 
perfect Abelian dominance ($\sigma_{\rm Abel} \simeq \sigma$), 
when the spatial size $La$ is larger than about $2$~fm.

\begin{figure}[h]
\centering
\includegraphics[height=4.5cm]{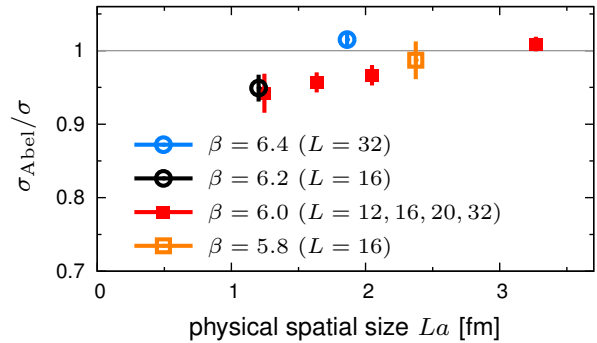}
\caption{
Physical spatial-size dependence of the ratio $\sigma_{\rm Abel}/\sigma$ \cite{SS15}. 
Perfect Abelian dominance ($\sigma_{\rm Abel} \simeq \sigma$) is found 
for larger lattices with $La \ge 2$~fm.
}
\end{figure}

\section{Quark Confinement in Baryons}

Second, we accurately measure the static three-quark (3Q) potential $V_{\rm 3Q}$ 
\cite{SS15} for more than 300 different patterns of 3Q systems 
with 1000-2000 gauge configurations 
in SU(3) lattice QCD on large physical-volume lattices with $La > 2{\rm fm}$: 
$(\beta, L^3 \times L_t)$=(5.8, $16^3 \times 32$) and (6.0, $20^3\times 32$).

\subsection{Accurate Measurement of Three-Quark Potential}

Like the Q$\bar{\rm Q}$ potential, 
the 3Q potential $V_{\rm 3Q}$ is obtained 
from the ``3Q Wilson loop'' $W_{\rm 3Q}$ (an extension of the Wilson loop 
for gauge-invariant static 3Q systems) defined in Ref.\cite{TS01}:
\begin{equation}
V_{\rm 3Q} = - \lim_{t \rightarrow \infty}\frac{1}{t}\ln \left\langle 
W_{\rm 3Q} \left[ U_\mu  \right] \right\rangle.
\end{equation}
%
We consider 101 and 211 different patterns 
of 3Q systems with  
$2000$ and $1000$ gauge configurations 
at $\beta$=5.8 and 6.0, respectively.
For the accurate calculation of the 3Q potential with finite $t$, 
we use the gauge-invariant smearing method \cite{TS01}, 
which enhances the ground-state component in the 3Q state in 
$\langle W_{\rm 3Q} \rangle$.

As the result, all the lattice QCD data of the 3Q potential $V_{\rm 3Q}$ 
is found to be fairly well reproduced by the Y-Ansatz \cite{SS15,TS01}, 
i.e., one-gluon-exchange Coulomb plus Y-type 
linear potential, 
\begin{eqnarray} 
V_{\rm 3Q} ({\bf r}_1, {\bf r}_2, {\bf r}_3)
= -\sum_{i<j} \frac{A_{\rm 3Q} }{|{\bf r}_i-{\bf r}_j|}
+\sigma_{\rm 3Q} L_{\rm min}+C_{\rm 3Q}
=- \frac{A_{\rm 3Q}}{R}
+\sigma_{\rm 3Q} L_{\rm min}+C_{\rm 3Q}, 
\label{eq:Y}
\end{eqnarray}
with $\sigma_{\rm 3Q}\simeq \sigma$ (Q$\bar{\rm Q}$ string tension). 
$L_{\rm min}$ is the minimal flux-tube length 
connecting the three quarks, located at ${\bf r}_1, {\bf r}_2$ and ${\bf r}_3$. 
We here introduce a convenient variable 
$1/R \equiv \sum_{i<j} 1/|{\bf r}_i-{\bf r}_j|$ \cite{B1013}.
The Y-Ansatz (\ref{eq:Y}) indicates 
the Y-shaped flux-tube formation in baryons, which is actually observed 
in lattice QCD calculations on the action density 
in the presence of three static quarks \cite{IBSS03}.

\subsection{Perfect Abelian dominance of quark confinement in baryons}

Next, we examine Abelian dominance of quark confinement in the 3Q system.
Like the Q$\bar{\rm Q}$ case, 
the MA-projected 3Q potential $V_{\rm 3Q}^{\rm Abel}$ (Abelian part)
is defined by the Abelian 3Q Wilson loop 
$W_{\rm 3Q} \left[u_\mu \right]$ in the MA gauge, 
\begin{eqnarray}
V_{\rm 3Q}^{\rm Abel} = - \lim_{t \rightarrow \infty}
\frac{1}{t}\ln \left\langle 
W_{\rm 3Q} \left[u_\mu \right] \right\rangle. 
\end{eqnarray}

Figure~3(a) shows the 3Q potential $V_{\rm 3Q}$ and 
its Abelian part $V_{\rm 3Q}^{\rm Abel}$ 
plotted against the total flux-tube length $L_{\rm min}$ 
in SU(3) lattice QCD at $\beta$=5.8 on $16^3 \times 32$ 
with 2,000 gauge configurations \cite{SS15}. 
The Abelian part $V_{\rm 3Q}^{\rm Abel}$ of the 3Q potential 
also takes the Y-Ansatz \cite{SS15}, 
\begin{eqnarray}
V_{\rm 3Q}^{\rm Abel}
= -\sum_{i<j} \frac{A_{\rm 3Q}^{\rm Abel} }{|{\bf r}_i-{\bf r}_j|}
+\sigma_{\rm 3Q}^{\rm Abel} L_{\rm min}+C_{\rm 3Q}^{\rm Abel}
= - \frac{A_{\rm 3Q}^{\rm Abel}}{R}
+\sigma_{\rm 3Q}^{\rm Abel} L_{\rm min}
+C_{\rm 3Q}^{\rm Abel},
\label{eq:C}
\end{eqnarray}
with $1/R \equiv \sum_{i<j} 1/|{\bf r}_i-{\bf r}_j|$.
At long distances, $V_{\rm 3Q}$ and $V_{\rm 3Q}^{\rm Abel}$ 
are almost single-valued functions of $L_{\rm min}$, although 
their multi-valued feature due to the $R$-dependence 
is more visible at short distances on finer lattices at $\beta=6.0$.
From Fig.3(a), we find {\it perfect Abelian dominance} also for 
the 3Q confinement force, i.e., 
$\sigma^{\rm Abel}_{\rm 3Q} \simeq \sigma_{\rm 3Q}\simeq \sigma$.

\begin{figure}[h]
\centering
\vspace{-0.5cm}
\includegraphics[height=4.8cm]{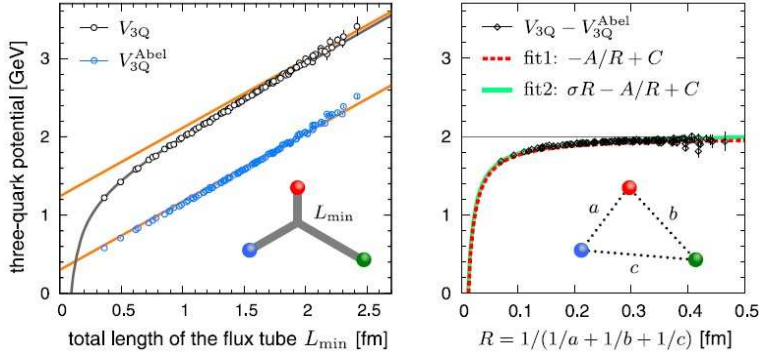}
\vspace{-0.1cm}
\caption{
(a) The 3Q potential $V_{\rm 3Q}$ (black) 
and its Abelian part $V_{\rm 3Q}^{\rm Abel}$ (blue) 
plotted against $L_{\rm min}$ 
in SU(3) lattice QCD at $\beta$=5.8 on $16^3 \times 32$.
We add the best-fit Y-Ansatz curve of the equilateral 3Q case 
for $V_{\rm 3Q}$ and $V_{\rm 3Q}^{\rm Abel}$, respectively. 
(b) $\Delta V_{\rm 3Q} \equiv V_{\rm 3Q} - V_{\rm 3Q}^{\rm Abel}$ 
plotted against $R$. 
$\Delta V_{\rm 3Q}$ can be fit with the pure Coulomb Ansatz (\ref{eq:C}) 
with no string tension, which indicates 
$\sigma_{\rm 3Q} \simeq \sigma_{\rm 3Q}^{\rm Abel}$.
These figures are taken from Ref.\cite{SS15}.
}
\end{figure}

To demonstrate $\sigma_{\rm 3Q}^{\rm Abel} \simeq \sigma_{\rm 3Q}$, 
we show in Fig.3(b) the difference 
$\Delta V_{\rm 3Q} \equiv V_{\rm 3Q}-V_{\rm 3Q}^{\rm Abel}$ 
plotted against $R$ \cite{SS15}, because,  
if $\sigma_{\rm 3Q}^{\rm Abel} = \sigma_{\rm 3Q}$, 
$\Delta V_{\rm 3Q}$ is well reproduced by the pure Coulomb Ansatz, 
\begin{equation}
\Delta V_{\rm 3Q} \equiv V_{\rm 3Q} - V_{\rm 3Q}^{\rm Abel}
= - \frac{\Delta A_{\rm 3Q}}{R} + \Delta C_{\rm 3Q},
\end{equation}
with $\Delta A_{\rm 3Q} \equiv A_{\rm 3Q}- A_{\rm 3Q}^{\rm Abel}$,
$\Delta C_{\rm 3Q} \equiv C_{\rm 3Q}-C_{\rm 3Q}^{\rm Abel}$ 
and $1/R \equiv \sum_{i<j} 1/|{\bf r}_i-{\bf r}_j|$.
In Fig.3(b), $\Delta V_{\rm 3Q}$ obeys 
a pure Coulomb form with no string tension, 
which is a clear evidence of {\it perfect Abelian dominance} 
of quark confinement in baryons: 
$\sigma_{\rm 3Q}^{\rm Abel} \simeq \sigma_{\rm 3Q}$ \cite{SS15}.

To summarize, from the analysis of the accurate lattice QCD data of 
$V(r)$, $V^{\rm Abel}(r)$, $V_{\rm 3Q}$ and $V_{\rm 3Q}^{\rm Abel}$ \cite{SS14,SS15}, 
we find {\it perfect Abelian dominance} of the string tension in 
Q$\bar{\rm Q}$ and 3Q potentials: 
$\sigma \simeq \sigma^{\rm Abel} 
\simeq \sigma_{\rm 3Q} \simeq \sigma_{\rm 3Q}^{\rm Abel}$.

\section{Monopole Dominance of Quark Confinement in Mesons and Baryons}

Finally, we accurately investigate monopole dominance 
of quark confinement for both Q$\bar{\rm Q}$ and 3Q systems 
in SU(3) lattice QCD 
at $\beta$=5.8 on $16^3 \times 32$ with 2,000 gauge configurations. 

\begin{figure}[h]
\centering
\includegraphics[height=5.5cm]{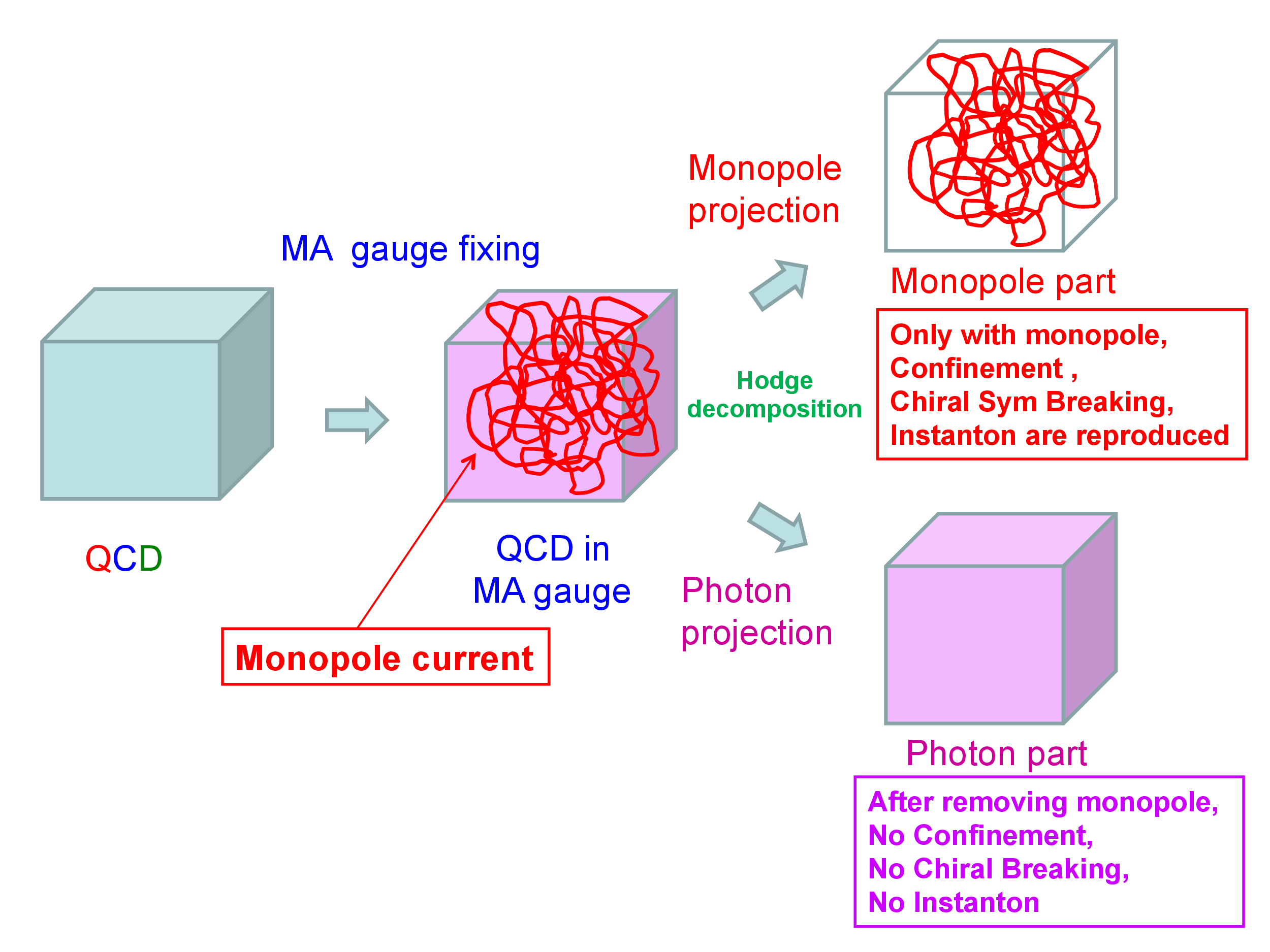}
\vspace{-0.3cm}
\caption{
A dual superconductor scenario from QCD in the MA gauge. 
In the MA gauge, infrared QCD becomes Abelian-like because of large 
off-diagonal gluon mass of about 1GeV \cite{AS99}, 
and monopole currents topologically appear \cite{KSW87}. 
By the Hodge decomposition, the QCD system in the MA gauge 
can be divided into the monopole part ($k_\mu \ne 0$, $j_\mu=0$) 
and the photon part ($j_\mu \ne 0$, $k_\mu=0$). 
}
\end{figure}

\subsection{Hodge Decomposition and Monopole Projection}

In the MA gauge, it is likely that only Abelian gluon component 
is essential for the long-distance QCD physics, 
and infrared QCD can be approximated by Abelian-projected QCD,  
as is indicated by perfect Abelian dominance of quark confinement. 

Abelian-projected QCD in the MA gauge has 
not only the color-electric current $j^\mu$ 
but also the color-magnetic monopole current $k^\mu$, 
which topologically appears. 
In the dual superconductor scenario, 
the monopole current $k^\mu$ is considered to play 
an essential role to quark confinement. 
By the Hodge decomposition, the Abelian-projected QCD system  
can be divided into the monopole part ($k_\mu \ne 0$, $j_\mu=0$) 
and the photon part ($j_\mu \ne 0$, $k_\mu=0$), 
as schematically illustrated in Fig.4. 
Then, the importance of the monopole current $k^\mu$ can be checked, 
using the Hodge decomposition.

In the lattice formalism, DeGrand and Toussaint performed 
the Hodge decomposition \cite{DT80}.
In lattice QCD, the Abelian gluon $\theta_\mu(s)=agA_\mu(s)$  
is the exponent in Abelian link-variable,  
\begin{eqnarray}
u_\mu(s)=e^{i \theta_\mu(s)}=e^{i \theta_\mu^3(s)T^3+i \theta_\mu^8(s)T^8}
={\rm diag}(e^{i\theta_\mu^{(1)}(s)},e^{i\theta_\mu^{(2)}(s)}, e^{i\theta_\mu^{(3)}(s)})
\in {\rm U(1)}^2,
\end{eqnarray}
with $\theta_\mu^{(i)}(s) \in [-\pi, \pi)$ ($i$=1,2,3), 
which is consistent with the continuum gluon $A_\mu$ as $a \rightarrow 0$.
The Abelian field strength $\theta_{\mu\nu}(s)=a^2gF_{\mu\nu}(s)$ 
is the exponent in the Abelian plaquette variable, 
\begin{eqnarray}
u_{\mu\nu}(s) 
=e^{i(\partial \wedge \theta)_{\mu\nu}(s)}
=e^{i \theta_{\mu\nu}(s)}
=e^{i \theta_{\mu\nu}^3(s)T^3+i \theta_{\mu\nu}^8(s)T^8}
={\rm diag}(e^{i\theta_{\mu\nu}^{(1)}(s)},e^{i\theta_{\mu\nu}^{(2)}(s)}, e^{i\theta_{\mu\nu}^{(3)}(s)}) \in {\rm U(1)}^2.~~~~
\end{eqnarray}
Here, $\theta_{\mu\nu}^{(i)}(s)$ 
is the principal value $\theta_{\mu\nu}^{(i)}(s) \in [-\pi, \pi)$ ($i$=1,2,3), which is 
${\rm U(1)}^2$-gauge invariant and consistent with 
the continuum Abelian field strength $F_{\mu\nu}$ as $a \rightarrow 0$ \cite{IS99}.
Then, $\theta_{\mu \nu}$ is written as  
\begin{eqnarray}
\theta_{\mu \nu}(s)=(\partial \wedge \theta)_{\mu\nu}(s)+2\pi n_{\mu\nu}(s),
\qquad n_{\mu\nu}^{(i)}(s) \in {\bf Z} ~~~(i=1,2,3),
\end{eqnarray}
where $n_{\mu\nu}(s)$ is U(1)$^2$ gauge-variant and 
corresponds to the singular Dirac string as $a\rightarrow0$ \cite{IS99}.
The electric current $j^\mu$ and the monopole current $k^\mu$ are 
derived from the Abelian field strength $\theta_{\mu\nu}$, 
\begin{eqnarray}
j_\nu \equiv \partial_\mu \theta_{\mu\nu}, \qquad
k_\nu \equiv \partial_\mu \tilde{\theta}_{\mu\nu}=2\pi \partial_\mu \tilde{n}_{\mu\nu}, \qquad
k_\nu^{(i)}=2\pi \partial_\mu \tilde{n}_{\mu\nu}^{(i)} \in 2\pi {\bf Z},
\end{eqnarray}
with the dual tensor 
$\tilde{\theta}_{\mu\nu}\equiv \frac12\epsilon_{\mu\nu\alpha\beta} \theta_{\alpha \beta}$.
The monopole part $\theta_\mu^{\rm Mo}$ and
the photon part $\theta_\mu^{\rm Ph}$ satisfy 
\begin{eqnarray}
\theta_{\mu \nu}^{\rm Mo}
&\equiv&(\partial \wedge \theta^{\rm Mo})_{\mu\nu} ~~({\rm mod} ~2\pi), 
\quad
\partial_\mu \theta_{\mu \nu}^{\rm Mo}=0, 
\quad
\partial_\mu \tilde{\theta}_{\mu \nu}^{\rm Mo}=k_\nu, 
\label{eq:M}\\
\theta_{\mu \nu}^{\rm Ph}
&\equiv&(\partial \wedge \theta^{\rm Ph})_{\mu\nu} ~~~({\rm mod} ~2\pi),
\quad
\partial_\mu \theta_{\mu \nu}^{\rm Ph}=j_\nu, 
\quad
\partial_\mu \tilde{\theta}_{\mu \nu}^{\rm Ph}=0. 
\label{eq:P}
\end{eqnarray}
From $\partial_\mu \tilde{\theta}_{\mu \nu}^{\rm Ph}=0$, one finds 
$\theta_{\mu \nu}^{\rm Ph}
=(\partial \wedge \theta^{\rm Ph})_{\mu\nu}$
and 
$\partial_\mu (\partial \wedge \theta^{\rm Ph})_{\mu\nu}
=\partial^2 \theta_\nu^{\rm Ph}-\partial_\nu(\partial_\mu\theta_\mu^{\rm Ph})=j_\nu$.
Taking the Landau gauge $\partial_\mu \theta_\mu^{\rm Ph}=0$,
the photon part $\theta_\nu^{\rm Ph}$ is derived from the electric current $j_\nu$, 
\begin{eqnarray}
\partial^2 \theta_\nu^{\rm Ph}=j_\nu, \quad
\theta_\nu^{\rm Ph}=\frac{1}{\partial^2} j_\nu, \quad
{\rm i.e.,} \quad \theta_\nu^{\rm Ph}(s)
=\sum_{s'} \langle s|\frac{1}{\partial^2}|s'\rangle j_\nu(s'), 
\end{eqnarray}
using the inverse d'Alembertian on the lattice \cite{IS99}.
The monopole part $\theta_\mu^{\rm Mo}(s)$ is obtained as 
$\theta_\mu^{\rm Mo}(s)=\theta_\mu(s)-\theta_\mu^{\rm Ph}(s)$.  
The monopole part $\theta_\mu^{\rm Mo}(s)$ and 
the photon part $\theta_\mu^{\rm Ph}(s)$ satisfy 
Eqs.(\ref{eq:M}) and (\ref{eq:P}) near the continuum with a small $a$ \cite{IS99}.

Using the monopole/photon link-variables, 
\begin{eqnarray}
u_\mu^{\rm Mo}(s) \equiv e^{i \theta_\mu^{\rm Mo}(s)} \in {\rm U(1)}^2, \qquad
u_\mu^{\rm Ph}(s) \equiv e^{i \theta_\mu^{\rm Ph}(s)} \in {\rm U(1)}^2,
\end{eqnarray}
monopole projection and photon projection are defined as follows:
\begin{itemize}
\item
Monopole projection (the monopole part) is defined by 
the replacement of $\{u_\mu(s)\} \rightarrow \{u_\mu^{\rm Mo}(s)\}$, 
which keeps the monopole current $k^\mu$ and eliminates the electric current $j^\mu$.
\item
Photon projection (the photon part)  is defined by 
the replacement of $\{u_\mu(s)\} \rightarrow \{u_\mu^{\rm Ph}(s)\}$, 
which keeps the electric current $j^\mu$ and 
eliminates the monopole current $k^\mu$. 
\end{itemize}
The dominant role of the monopole part is called ``monopole dominance'', 
and monopole dominance has been investigated for quark confinement 
in lattice QCD \cite{SNW94}.

\subsection{Monopole Dominance of Confinement for Quark-Antiquark and 3Q Systems}

The monopole part $V_{\rm Mo}(r)$ and the photon part $V_{\rm Ph}(r)$ 
of the Q$\bar{\rm Q}$ potential 
are defined by the monopole/photon-projected Wilson loop, 
$W_{r \times t}[u_\mu^{\rm Mo}]$ and
$W_{r \times t}[u_\mu^{\rm Ph}]$,
\begin{eqnarray}
V_{\rm Mo}(r)=-\lim_{t \rightarrow \infty} 
\frac{1}{t}{\rm ln} \langle W_{r \times t}[u_\mu^{\rm Mo}] \rangle, \qquad
V_{\rm Ph}(r)=-\lim_{t \rightarrow \infty} 
\frac{1}{t}{\rm ln} \langle W_{r \times t}[u_\mu^{\rm Ph}] \rangle.
\end{eqnarray}
Figure~5(a) shows the lattice QCD result for 
the static Q$\bar{\rm Q}$ potential $V(r)$ in SU(3) QCD, 
$V_{\rm Abel}(r)$ in Abelian-projected QCD,  
$V_{\rm Mo}(r)$ in the monopole part, and 
$V_{\rm Ph}(r)$ in the photon part. 
While the photon part has almost no confining force, 
the monopole part almost keeps the confining force.  
Thus, monopole dominance is found for quark confinement in the Q$\bar{\rm Q}$ system.

\begin{figure}[h]
\centering
\includegraphics[height=5.3cm]{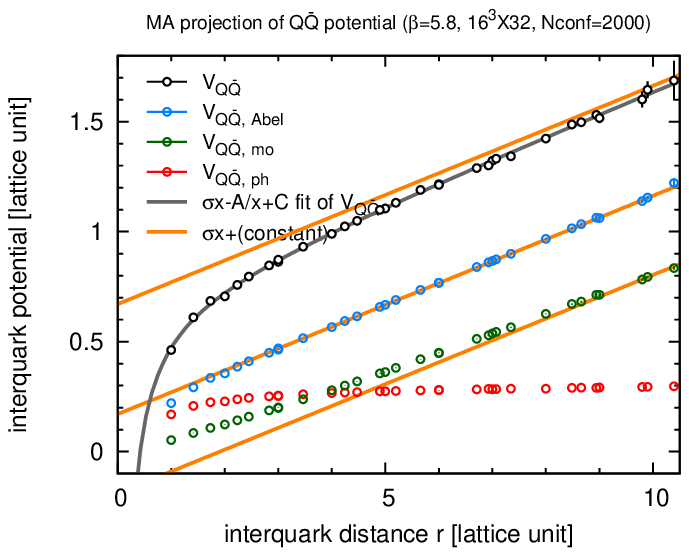}
\hspace{0.3cm}
\includegraphics[height=6cm]{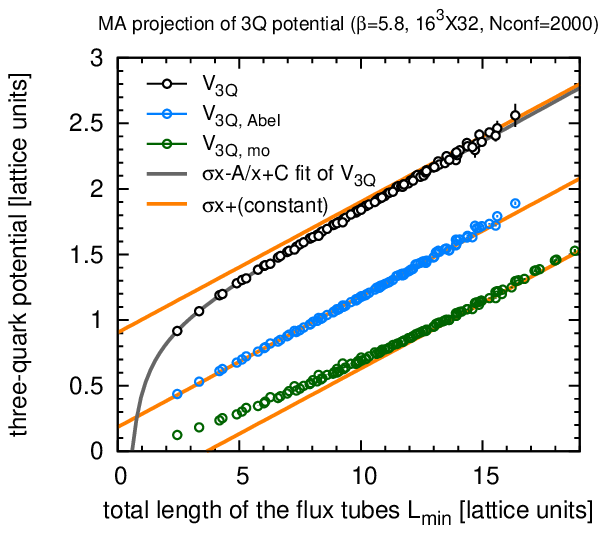}
\caption{
(a) The lattice QCD result for 
the static Q$\bar{\rm Q}$ potential $V(r)$ (black) in SU(3) QCD, 
$V_{\rm Abel}(r)$ (blue) in Abelian-projected QCD,  
$V_{\rm Mo}(r)$ (green) in the monopole part, and 
$V_{\rm Ph}(r)$ (red) in the photon part. 
(b) The lattice QCD result for the 3Q potential $V_{\rm 3Q}$ (black) in SU(3) QCD, 
$V_{\rm 3Q}^{\rm Abel}$ (blue) in Abelian-projected QCD, 
and $V_{\rm 3Q}^{\rm Mo}$ (green) in the monopole part, 
plotted against $L_{\rm min}$.
}
\end{figure}

Similarly, the monopole part $V_{\rm 3Q}^{\rm Mo}(r)$ of 
the 3Q potential is defined by  
\begin{equation}
V_{\rm 3Q}^{\rm Mo} = - \lim_{t \rightarrow \infty}\frac{1}{t}\ln \left\langle 
W_{\rm 3Q} \left[ u_\mu^{\rm Mo}  \right] \right\rangle. 
\end{equation}
Figure~5(b) shows the lattice QCD result for the 3Q potential 
$V_{\rm 3Q}$ in SU(3) QCD, 
$V_{\rm 3Q}^{\rm Abel}$ in Abelian-projected QCD, 
and $V_{\rm 3Q}^{\rm Mo}$ in the monopole part, 
plotted against $L_{\rm min}$.
Monopole dominance is found also for quark confinement in the 3Q system.

The string tension $\sigma_{\rm Mo}$ in the monopole part is estimated  
from the lattice QCD data in Fig.5, 
and monopole dominance is estimated as 
$\sigma_{\rm Mo} \simeq  0.92 \sigma$ 
for the string tension in Q$\bar{\rm Q}$ and 3Q systems. 

\section{Summary and concluding remarks}

We have investigated Abelian dominance and monopole dominance of quark confinement 
for Q$\bar{\rm Q}$ and 3Q systems in SU(3) quenched lattice QCD 
in the MA gauge. 
For large physical-volume lattices with $La \ge 2{\rm fm}$, 
we have found {\it perfect Abelian dominance} of the string tension 
for both Q$\bar{\rm Q}$ and 3Q systems: 
$\sigma \simeq \sigma^{\rm Abel} 
\simeq \sigma_{\rm 3Q} \simeq \sigma_{\rm 3Q}^{\rm Abel}$.
We have found monopole dominance of the string tension 
for both Q$\bar{\rm Q}$ and 3Q systems, 
and have estimated $\sigma_{\rm Mo} \simeq 0.92 \sigma$. 

\begin{acknowledgments}
H.S. is supported in part by the Grants-in-Aid  for Scientific Research 
[15K05076] from Japan Society for the Promotion of Science.
The lattice QCD calculations were done on NEC-SX8R 
at Osaka University.
\end{acknowledgments}

\end{document}